\documentclass[A4paper,twocolumn]{article}
\usepackage{amssymb}

\usepackage{graphicx}
\usepackage{amsmath}

\newtheorem{lemma}{Lemma}
\newtheorem{theorem}{Theorem}

\begin{document}

\title{Matched witness for multipartite entanglement}
\author{Xiao-yu Chen$^a$\thanks{%
Email:xychen@zjgsu.edu.cn}, Zhu-an Xu$^b$ \\
{\small {$^a$College of Information and Electronic Engineering, Zhejiang
Gongshang University, Hangzhou, Zhejiang 310018, China }}\\
{\small {$^b$Department of Physics, Zhejiang University, Hangzhou, Zhejiang
310027, China }}}
\date{}
\maketitle

\begin{abstract}
We transform the way of finding entanglement criterion into two steps: to
obtain necessary criterion of separability by maximizing an algebra function
for a set of characteristic variables of the witness operator and the given
number of partitions, then to obtain the sufficient criterion by minimizing
an algebra function with respect to the characteristic variables for a given
quantum state. Our method avoids the semi-definite program calculation in
the witness operator entanglement detection. The necessary and sufficient
criterion of separability for the three qubit X shaped state is given to
illustrate the procedure of finding the criterion. We give the necessary and
sufficient criteria of the three partite and full separabilities for the
four qubit noisy GHZ state and the four qubit noisy cluster state.

PACS number(s): 03.67.Mn; 03.65.Ud

Keywords: Multipartite entanglement; Separable criterion; Cluster state; GHZ
state \newline
\end{abstract}

\section{Introduction}

Quantum entanglement is considered as the central resource in quantum
communication and quantum computation. It is a special form of bipartite or
multipartite quantum superposition, and intrinsically different from any
classical system for its structures and properties. Closely related to
entanglement is the separable state. A separable state is the state which
can be written as the probability mixture of product states\cite{Werner}. An
entangled state is simply not separable. To determine a given state is
entangled or not is still very difficult in general now. Thus entanglement
criteria are desirable. Many progresses have been achieved for the criteria
of entanglement or separability. Among them are the Peres-Horodecki
criterion \cite{Peres} \cite{Horodecki}, the computable cross norm \cite
{Rodolph}or realignment criterion \cite{KChen} , the entropy criterion, the
uncertainty criterion \cite{Guhne2004} and so on \cite{Doherty} \cite{Li}.
All these criteria are necessary conditions for separability. Violation of
them means entanglement. The criteria are mainly applied to bipartite
states, the criteria for multipartite entanglement are sporadic \cite
{GuhneNJP,GuhnePRL}. One the other hand, entanglement witness is also a mean
for entanglement detection \cite{Terhal}, especially friendly for
experiment. In this paper, we will start from the original definition of
separable state, combine with entanglement witness method to derive
multipartite entanglement criteria. We find that the process in finding the
necessary criteria may also give rise to the sufficient criteria of
separability.

\section{Entanglement witness in characteristic form}

A multipartite state $\rho $ is separable when it can be written as \cite
{Werner}
\begin{equation}
\rho =\sum_ip_i\rho _i^{A_1}\otimes \rho _i^{A_2}\otimes \ldots \otimes \rho
_i^{A_N}.  \label{we0}
\end{equation}
where $\rho _i^{A_j}$ is the state (it is always possible to assume it to be
pure) of $A_j$ part, $p_i$ form a probability distribution. The method of
partition of the system into $N$ parts may change with the index $i$ to cope
with the concepts of biseparable, $k-$separable and fully separable. In the
following, we will consider the three qubit system. Extending to
multipartite qudit system is straightforward. For the problem of full
separability of a three qubit state $\rho $, its characteristic function is
\begin{equation}
R_{ijk}=tr[\rho (\sigma _i\otimes \sigma _j\otimes \sigma _k)],  \label{we1}
\end{equation}
where $i,j,k=0,1,2,3;$ $\sigma _i$ are Pauli matrices for $i=1,2,3$ and $%
\sigma _0=I_2$ is the $2\times 2$ identity matrix. Let the characteristic
function of an entanglement witness $W$ be $-M_{ijk}$(We call $M_{ijk}$
characteristic variables). We set $M_{ijk}$ to be zero when $R_{ijk}$ is
zero for simplicity. A state $\rho $ is entangled when
\begin{equation}
-tr(\rho W)=8\sum_{ijk}R_{ijk}M_{ijk}>0,  \label{we2}
\end{equation}
while for all fully separable states $\rho ^s$ we have
\begin{equation}
-tr(\rho ^sW)=8\sum_{ijk}R_{ijk}^sM_{ijk}\leq 0.  \label{we3}
\end{equation}
The problem then is reduced to find a series of parameters $M_{ijk}$. For a
three particle fully separable state $\rho ,$ its characteristic function is
$R_{ijk}^s=\sum_lp_lx_{il}y_{jl}z_{kl},$ where $x_{il}=tr(\rho
_l^{A_1}\sigma _i),$ with $x_{0l}=1,$ unit (Bloch) vector $\mathbf{x}_l%
\mathbf{=}(x_{1l},x_{2l},x_{3l})$ (since $\rho _l^{A_1}$is pure) and similar
for $y_{jl},z_{kl}$. Notice that for any state $R_{000}=1,$ then $%
\sum_{ijk}R_{ijk}^sM_{ijk}=\sum_lp_lx_{il}y_{jl}z_{kl}M_{ijk}=\sum_l^{\prime
}p_lx_{il}y_{jl}z_{kl}M_{ijk}+M_{000}\leq 0,$ where the prime in the
summation means that the $i=j=k=0$ term is removed from the summation. We
define $M_{000}\equiv -\max_{\mathbf{x,y,z}}\sum_{ijk}^{\prime
}M_{ijk}x_iy_jz_k$ with $x_0=y_0=z_0=1,$ $\left| \mathbf{x}\right| =\left|
\mathbf{y}\right| =\left| \mathbf{z}\right| =1$. Such a definition leads to
a legal entanglement witness in the sense that its keeps $tr(\rho W)\geq 0$
for all fully separable states. For given parameters $M_{ijk}$ with $M_{000}$
yet to be determined, the maximization $\max_{\mathbf{x,y,z}%
}\sum_{ijk}^{\prime }M_{ijk}x_iy_jz_k$ can be carried out analytically or
numerically. Denote vectors $\mathbf{M=}(M_{001,}M_{002},M_{003},\ldots
,M_{333})$ and $\mathbf{R=}(R_{001,}R_{002},R_{003},\ldots ,R_{333}),$ then
the necessary condition of full separability is
\begin{equation}
\mathbf{M\cdot R\leq -}M_{000}.  \label{we4}
\end{equation}
Violation of it implies entanglement. To find the matched witness, we
calculate the following minimization,
\begin{equation}
p=\min_{\mathbf{M}}\frac{-M_{000}}{\mathbf{M\cdot R}}.  \label{we5}
\end{equation}
Keep in mind that $\mathbf{M}$ is so chosen such that $\mathbf{M\cdot R}$ is
positive, $M_{000}$ is negative. $1-p$ is the white noise tolerance of an
entangled state $\rho .$ If the state is fully separable, then $p$ $\geq 1.$

For the problem of biseparability of a three qubit state, we first consider
the characteristic function of a pure two qubit state $\left| \psi
\right\rangle =\sum_{m,n=0}^1\alpha _{mn}\left| mn\right\rangle $. The
characteristic function of $\left| \psi \right\rangle $ is denoted as $%
T_{ij}=\left\langle \psi \right| \sigma _i\otimes \sigma _j\left| \psi
\right\rangle .$ A state $\rho ^{bs}$ is biseparable if
\begin{eqnarray}
\rho ^{bs} &=&\sum_l(p_{1l}\rho _l^{A_1}\otimes \rho _l^{A_2A_3}+p_{2l}\rho
_l^{A_2}\otimes \rho _l^{A_1A_3}  \nonumber \\
&&+p_{3l}\rho _l^{A_3}\otimes \rho _l^{A_1A_2}),  \label{we6}
\end{eqnarray}
where $\rho _l^{A_mA_n}$ is a pure two qubit state. The genuine entanglement
witness $W$ with characteristic function $-M_{ijk}$ should fulfill the
condition $-tr(\rho ^{bs}W)=8\sum_{ijk}R_{ijk}^{bs}M_{ijk}\leq 0$ for any
biseparable state $\rho ^{bs}$. The characteristic function of $\rho ^{bs}$
is $R_{ijk}^{bs}=$ $\sum_l%
\sum_{ijk}(p_{1l}x_{il}T_{jk}^{(23)}+p_{2l}y_{il}T_{jk}^{(13)}+p_{3l}z_{il}T_{jk}^{(12)}).
$ Let $M_x=\max_{\mathbf{x,T}^{(23)}}\sum_{ijk}^{\prime
}M_{ijk}x_iT_{jk}^{(23)},$ and $M_{000}=-\max \{M_x,M_y,M_z\},$ where $%
M_y,M_z$ are defined similarly as $M_x.$ The necessary condition of
biseparability is still described by inequality (\ref{we4}), with a new
definition of $M_{000}.$ The noise tolerance of a genuine entangled state is
$1-p$ too, with a different $p$ from a different definition of $M_{000}.$

\section{Three qubit X shaped states\textit{\ }}

The density matrix of a three qubit $X$ shaped state is a $8\times 8$ matrix
with diagonal entries, anti-diagonal entries and all the other entries are
zero. The possible nonzero entries are denoted as $\rho _{00},\rho
_{11},\ldots ,\rho _{77}$ for diagonal elements and $\rho _{07},\rho
_{16},\ldots ,\rho _{70}$ for anti-diagonal elements. The decimal subscript $%
l$ is equivalent to the three bit binary string $l_1l_2l_3$ such that $%
l=4l_1+2l_2+l_3$. Hence we may write $\rho _{16}$ as $\rho _{001,110}$ in
three bit binary subscripts. For an three qubit $X$ state, the nonzero
elements of the characteristic function are $R_{ijk}$ with the subscripts $%
i,j,k=0,3$ or $i,j,k=1,2$. The unit vector $\mathbf{x}$ can be written as $%
\mathbf{x=}(\sin \theta _1\cos \varphi _1,\sin \theta _1\sin \varphi _1,\cos
\theta _1)$ with $\theta _1\in [0,\pi ],$ $\varphi _1\in [0,2\pi ].$
Similarly, $\mathbf{y,z}$ are expressed with $\theta _2,\varphi _2$ and $%
\theta _3,\varphi _3,$ respectively. Then $-M_{000}=\max_{\mathbf{\theta
,\varphi }}F(\mathbf{\theta ,\varphi }),$ with
\[
F(\mathbf{\theta ,\varphi })=f(\mathbf{\theta })+s_1s_2s_3g(\mathbf{\varphi }%
),
\]
and $\mathbf{\theta }=(\theta _1,\theta _2,\theta _3)\mathbf{,\varphi =}%
(\varphi _1,\varphi _2,\varphi _3),$ where $f(\mathbf{\theta }%
)=M_{003}c_3+M_{030}c_2+M_{300}c_1+M_{033}c_2c_3$ $%
+M_{303}c_1c_3+M_{330}c_1c_2$ $+M_{333}c_3c_2c_1,$ we denote $\cos \theta
_i=c_i,\sin \theta _i=s_i$ for short, and
\begin{eqnarray}
g(\mathbf{\varphi }) &=&M_{111}c_1^{\prime }c_2^{\prime }c_3^{\prime
}+M_{122}c_1^{\prime }s_2^{\prime }s_3^{\prime }  \nonumber \\
&&+M_{212}s_1^{\prime }c_2^{\prime }s_3^{\prime }+M_{221}s_1^{\prime
}s_2^{\prime }c_3^{\prime }.  \label{we8}
\end{eqnarray}
for the case of real anti-diagonal entries of an $X$ shaped state. Where $%
c_i^{\prime }=\cos \varphi _i,s_i^{\prime }=\sin \varphi _i.$

\begin{lemma}
Denote the maximum of $g(\mathbf{\varphi })$ in (\ref{we8}) with respect to $%
\varphi $ as $g_m,$ then
\begin{equation}
g_m=\left\{
\begin{array}{l}
\sqrt{\frac{(\delta \alpha +\beta \gamma )(\delta \beta +\alpha \gamma
)(\delta \gamma +\alpha \beta )}{\delta \alpha \beta \gamma }},\text{ for }%
\delta \alpha \beta \gamma >0\ \text{and }q\geq 0; \\
\text{ }\max \{\left| M_{111}\right| ,\left| M_{122}\right| ,\left|
M_{212}\right| ,\left| M_{221}\right| \},\text{otherwise.}
\end{array}
\right.  \label{we9}
\end{equation}
Where $(\delta ,\alpha ,\beta ,\gamma )=\frac
14(M_{111},M_{122},M_{212},M_{221})\Gamma ,$ here $\Gamma $ is a $4\times 4$
matrix with all of its diagonal entries being $-1$ and off-diagonal entries
being $+1.$ $q=q_0q_1q_2q_3$ with $(q_0,q_1,q_2,q_3)=(\alpha \beta \gamma
,\delta \beta \gamma ,\delta \alpha \gamma ,\delta \alpha \beta )\Gamma .$
\end{lemma}

The proof can be found in Appendix.

\begin{theorem}
An $X$ shaped state with identical diagonal entries and real anti-diagonal
entries is separable iff
\[
R\leq 1,
\]
where
\begin{equation}
R=\left\{
\begin{array}{l}
\sqrt{\frac{(R_0R_1+R_2R_3)(R_0R_2+R_1R_3)(R_0R_3+R_1R_2)}{R_0R_1R_2R_3}},
\\
\text{ \ \qquad \qquad \qquad for }Q>0\text{ and }r\geq 0; \\
8\max \{\left| \rho _{07}\right| ,\left| \rho _{16}\right| ,\left| \rho
_{25}\right| ,\left| \rho _{34}\right| \},\text{ otherwise.}
\end{array}
\right.  \label{we9a}
\end{equation}
with $R_0=R_{111},$ $R_1=R_{122},$ $R_2=R_{212},$ $R_3=R_{221}$ and $%
Q=R_0R_1R_2R_3.$ Here $r=r_0r_1r_2r_3$ and vector $%
(r_0,r_1,r_2,r_3)=(R_1R_2R_3,R_0R_2R_3,R_0R_1R_3,R_0R_1R_2)\Gamma .$
\end{theorem}

Proof: ``Only if'': At present case, we have $-M_{000}=g_m$ by assuming all
the $M_{ijk}$ be $0$ except $M_{000}$ and $M_{111},M_{122},M_{212},M_{221}.$

In (\ref{we9}), if $g_m=\max \{\left| M_{111}\right| ,\left| M_{122}\right|
,\left| M_{212}\right| ,\left| M_{221}\right| \},$ let $\left|
M_{111}\right| $ is the biggest w.l.o.g. The necessary condition (\ref{we4})
is
\begin{equation}
L\equiv M_{111}R_0+M_{122}R_1+M_{212}R_2+M_{221}R_3\leq \left|
M_{111}\right| .  \label{we9b}
\end{equation}
We may choose the signs of $M_{111},M_{122},M_{212},M_{221}$ match the signs
of $R_i$ $(i=0,\ldots ,3)$ to make $L$ larger. If $Q<0,$ then one or three $%
R_i$ are negative. Let $R_0\ $be negative and the other $R_i$ be positive
for definite. We may further require $\left| M_{122}\right| =\left|
M_{212}\right| =\left| M_{221}\right| =\left| M_{111}\right| $ in order to
make the left side of (\ref{we9b}) even larger. This is possible if we
choose $\alpha =\beta =\gamma =0.$ For we obtain $\delta \left(
-R_0+R_1+R_2+R_3\right) \leq \left| \delta \right| $. Thus $8\left| \rho
_{07}\right| \leq 1,$ and there are similarly inequalities for the other
anti-diagonal entries. We then have $8\max \{\left| \rho _{07}\right|
,\left| \rho _{16}\right| ,\left| \rho _{25}\right| ,\left| \rho
_{34}\right| \}\leq 1.$ When $\delta \alpha \beta \gamma <0,$ let $\left|
\rho _{07}\right| $ be the largest one among $\{\left| \rho _{07}\right|
,\left| \rho _{16}\right| ,\left| \rho _{25}\right| ,\left| \rho
_{34}\right| \}.$ If $\rho _{07}>0,$ we may choose $\delta <0$ and $\alpha
,\beta ,\gamma >0,$ we have $\frac L{g_m}-8\rho _{07}\leq 0,$the equality is
achieved when $\left| \delta \right| \rightarrow \infty .$ If $\rho _{07}<0,$
we may choose $\delta ,\alpha ,\beta >0$ and $\gamma <0,$we have $\frac
L{g_m}+8\rho _{07}\leq 0,$the equality is achieved when $\delta \rightarrow
\infty .$ Thus we have proved that $R=\max_{\mathbf{M}}\frac L{g_m}=8\max
\{\left| \rho _{07}\right| ,\left| \rho _{16}\right| ,\left| \rho
_{25}\right| ,\left| \rho _{34}\right| \}$ for the cases of $Q<0$ or $\delta
\alpha \beta \gamma <0.$

If $g_m=\sqrt{(\delta \alpha +\beta \gamma )(\delta \beta +\alpha \gamma
)(\delta \gamma +\alpha \beta )/(\delta \alpha \beta \gamma )},$ we may
rewrite equation (\ref{we5}) as $p=\min_{\delta ,\alpha ,\beta ,\gamma }%
\frac{g_m}L,$ where $L=8(-\delta \rho _{07}+\gamma \rho _{16}+\beta \rho
_{25}+\alpha \rho _{34})$ is assumed to be positive. Then $\frac{\partial p}{%
\partial \delta }=0$ leads to $\frac{\partial \ln g_m}{\partial \delta }=%
\frac{\partial \ln L}{\partial \delta },$ which is $-8\rho _{07}=\frac
L2\left( \frac \alpha {\delta \alpha +\beta \gamma }+\frac \beta {\delta
\beta +\alpha \gamma }+\frac \gamma {\delta \gamma +\alpha \beta }-\frac
1\delta \right) ,$similarly, $\rho _{16},\rho _{25},\rho _{34}$ can also be
obtained. Then after some algebra we have $R_0=GLq_1q_2q_3,$ $%
R_1=GLq_0q_2q_3,$ $R_2=GLq_0q_1q_3,$ $R_3=GLq_0q_1q_2,$ with $G=\frac
1{8g_m^2(\delta \alpha \beta \gamma )^2}.$ The first line of the right hand
side of (\ref{we9a}) can be written as
\begin{equation}
R=GL\sqrt{(q_0q_1+q_2q_3)(q_0q_2+q_1q_3)(q_0q_3+q_1q_2)},  \label{we9d}
\end{equation}
Note that $q_0q_1+q_2q_3=4\delta \alpha \beta \gamma (\delta \alpha +\beta
\gamma ),$ we arrive at $R=\frac L{g_m}.$ Thus the noise tolerance of the
state is $p=\frac 1R.$ The state is fully separable when $R\leq 1.$ The
condition $Q>0$ guarantees the validation of the first line of (\ref{we9a}).
Since $Q=G^4L^4q^3,$ so $Q\geq 0$ is equivalent to $q\geq 0.$ Also we have $%
r=(4q^2G^3L^3)^4(\delta \alpha \beta \gamma )^3,$ hence $r>0$ is equivalent
to $\delta \alpha \beta \gamma >0.$

``If'': In the case of $Q>0$ (and $r>0$), that the state is fully separable
when $R\leq 1$is shown in \cite{GuhnePLA} \cite{Kay} \cite{ChenQIP}. In the
coordinate of $(\rho _{07},\rho _{16},\rho _{25},\rho _{34})$, the shape of
the fully separable state set is as follow: consider a four dimensional
hypercube centered at original and with side length $1/4,$ the $16$ vertices
are located at $\left| \rho _{07}\right| =\left| \rho _{16}\right| $ $%
=\left| \rho _{25}\right| =\left| \rho _{34}\right| $ $=\frac 18.$ The
vertex $(\rho _{07},\rho _{16},\rho _{25},\rho _{34})=(\frac 18,\frac
18,\frac 18,\frac 18)$ corresponds to the state $\frac 18(III+\sigma
_1\sigma _1\sigma _1),$ which is fully separable. The vertices with two or
four anti-diagonal elements being $-\frac 18$ corresponds to states which
are local equivalent to the fully separable state $\frac 18(III+\sigma
_1\sigma _1\sigma _1).$ The vertex $(\rho _{07},\rho _{16},\rho _{25},\rho
_{34})=(\frac 18,\frac 18,\frac 18,-\frac 18)$ corresponds to an entangled
state. The corner containing this vertex is cut by the hyper surface $R=1$
where $R$ is the first line of (\ref{we9a}). Similarly, the other $7$
corners with vertices which have odd number of $-\frac 18$ coordinate
components are also cut. Thus the fully separable state set is a four
dimensional hypercube with $8$ of its corners cut. The $8$ vertex states
with even number of $-\frac 18$ coordinate components and the states in the
cut surfaces are fully separable. This guarantees all the states correspond
to the inner and surface points in the cut hypercube is fully separable.$%
\blacksquare $

Denote $F_1(\mathbf{\theta })=\max_{\mathbf{\varphi }}F(\mathbf{\theta
,\varphi })=f(\mathbf{\theta })+g_ms_1s_2s_3,$ and the maximum of $F_1(%
\mathbf{\theta })$ with respect to $\theta _3$ as $F_2\left( \theta
_1,\theta _2\right) ,$ then
\begin{equation}
F_2\left( \theta _1,\theta _2\right) =a+bc_2+\sqrt{(c+dc_2)^2+e^2s_2^2}.
\label{we12}
\end{equation}
Where $a=M_{300}c_1,$ $b=M_{030}+M_{330}c_1,$ $c=M_{003}+M_{303}c_1,$ $%
d=M_{033}+M_{333}c_1,$ $e=g_ms_1.$

We consider two cases. Case (i), let us assume $e^2=d^2-c^2,$ which requires
$M_{033}=M_{330},$ $M_{030}=M_{333}$ (or $M_{033}=-M_{330},$ $%
M_{030}=-M_{333}$), $M_{033}^2=M_{030}^2+g_m^2.$ Then (\ref{we12}) is $%
F_2\left( \theta _1,\theta _2\right) =a+bc_2+\left| d+cc_2\right| .$
Assuming $d+cc_2<0,$ then $F_2\left( \theta _1,\theta _2\right)
=a+bc_2-d-cc_2.$ Let $b=c,$ that is $M_{030}=M_{003},M_{330}=M_{303},$ then $%
F_2\left( \theta _1,\theta _2\right) =a-d.$ If we further assume $%
M_{300}=M_{333},$ then $F_2\left( \theta _1,\theta _2\right) =-M_{033}$ is a
constant. Hence if we choose the set of parameters as $%
M_{030}=M_{333}=M_{300}=M_{003}=\sin \eta ,$ $g_m=\cos \eta
,M_{033}=M_{330}=M_{303}=-1,$ then all the assumptions $e^2=d^2-c^2,$ $%
b=c,d+cc_2<0$ can be fulfilled. The necessary condition of full separability
(\ref{we4}) reads
\begin{equation}
4\sin \eta (\rho _{00}-\rho _{77})+R\cos \eta \leq 4(\rho _{00}+\rho _{77})
\label{we13}
\end{equation}
for any $\eta .$ We have used $R_{003}+R_{030}+R_{300}+R_{333}=4(\rho
_{00}-\rho _{77})$ and $1+R_{033}+R_{303}+R_{330}=4(\rho _{00}+\rho _{77}).$
The necessary condition can be refined to $\sqrt{16(\rho _{00}-\rho
_{77})^2+R^2}\leq 4(\rho _{00}+\rho _{77})\ $, which is $\frac 18R\leq \sqrt{%
\rho _{00}\rho _{77}}.$ Similarly, we obtain $\frac 18R\leq \sqrt{\rho
_{33}\rho _{44}}\ $by choosing $M_{033}=-M_{330},$ $M_{030}=-M_{333}$
instead. If we properly choosing the parameters such that $e^2=d^2-c^2,$ $%
b=-c,d+cc_2>0$, then we arrive at the necessary conditions of $\frac 18R\leq
\sqrt{\rho _{22}\rho _{55}}$ and $\frac 18R\leq \sqrt{\rho _{11}\rho _{66}}$%
. Hence we have the necessary condition
\begin{equation}
\frac 18R\leq \min_{i=0,\ldots ,3}\sqrt{\rho _{ii}\rho _{7-i,7-i}}.
\label{we14}
\end{equation}

Case (ii), let us assume $M_{033}=M_{300}\equiv M_1,$ $M_{303}=M_{030}\equiv
M_2,$ $M_{003}=M_{330}\equiv M_3,$ $M_{333}\equiv M_0$ and $g_m=\sqrt[4]{%
m_0m_1m_2m_3},$ with the vector $(m_0,m_1,m_2,m_3)$ $=-\frac
12H(M_0,M_2,M_1,M_3)$. Here $H$ is the $4\times 4$ Hadamard matrix. Let $M_0$
be negative and $-M_0$ be sufficiently large. Then we can prove that
\begin{equation}
M_{000}=M_{333}=M_0,  \label{we14a}
\end{equation}
see Appendix for details. Hence the necessary condition of full separability
is
\begin{equation}
\frac 18g_mR\leq m_0\rho _{00}+m_1\rho _{33}+m_2\rho _{55}+m_3\rho _{66},
\label{we15}
\end{equation}
which is true for all possible choices of $m_i$ ($i=0,\ldots ,3$). Notice
that $m_0\rho _{00}+m_1\rho _{33}+m_2\rho _{55}+m_3\rho _{66}\geq 4\sqrt[4]{%
m_0m_1m_2m_3}\sqrt[4]{\rho _{00}\rho _{33}\rho _{55}\rho _{66}},$ the
identity is achieved when
\begin{equation}
m_0\rho _{00}=m_1\rho _{33}=m_2\rho _{55}=m_3\rho _{66}.  \label{we16}
\end{equation}
Thus the condition (\ref{we15}) can be refined as $\sqrt[4]{\rho _{00}\rho
_{33}\rho _{55}\rho _{66}}\geq \frac 18R.$ Similarly we have $\sqrt[4]{\rho
_{11}\rho _{22}\rho _{44}\rho _{77}}\geq \frac 18R.$ Hence the necessary
condition of full separability is refined as
\begin{equation}
\min (\sqrt[4]{\rho _{00}\rho _{33}\rho _{55}\rho _{66}},\sqrt[4]{\rho
_{11}\rho _{22}\rho _{44}\rho _{77}})\geq \frac 18R.  \label{we17}
\end{equation}
Here the role of matched witness is clearly shown by (\ref{we16}).

\begin{theorem}
A three qubit $X$ shaped state with real anti-diagonal entries is fully
separable iff
\begin{equation}
\min_{i=0,\ldots ,3}(\sqrt[4]{\rho _{00}\rho _{33}\rho _{55}\rho _{66}},%
\sqrt[4]{\rho _{11}\rho _{22}\rho _{44}\rho _{77}},\sqrt{\rho _{ii}\rho
_{7-i,7-i}})\geq \frac 18R.  \label{we18}
\end{equation}
\end{theorem}

Proof: The ``only if'' comes from (\ref{we14}) and (\ref{we17}). For the
``if '' part, consider the operator identity $A_1A_2A_3+B_1B_2B_3=\frac
14[(A_1+B_1)(A_2+B_2)(A_3+B_3)$ $%
+(A_1+B_1)(A_2-B_2)(A_3-B_3)+(A_1-B_1)(A_2+B_2)(A_3-B_3)$ $%
+(A_1-B_1)(A_2-B_2)(A_3+B_3)]$ , let $A_i=I+\cos \theta _i\sigma _3,$ $%
B_i=\sin \theta _i(\cos \varphi _i\sigma _1+\sin \varphi _i\sigma _2),$ then
the state $\frac 18\left( A_1A_2A_3+B_1B_2B_3\right) $ is fully separable
for any $\theta _i$ and $\varphi _i.$ Let $B_i^{\prime }=\sin \theta _i(\cos
\varphi _i\sigma _1-\sin \varphi _i\sigma _2),$ then the state $\varrho
=\frac 1{16}\left( 2A_1A_2A_3+B_1B_2B_3+B_1^{\prime }B_2^{\prime
}B_3^{\prime }\right) $ is a fully separable $X$ shaped state with real
anti-diagonal entries. The anti-diagonal entries have been treated in
Theorem 1, we have $R=\left| \sin \theta _1\sin \theta _2\sin \theta
_3\right| $ for the fully separable state $\varrho .$ In the case that all
the terms in the bracket of the left hand side of (\ref{we18}) are equal, it
is always possible to choose proper $\theta _i$ such that the state $\rho $
in (\ref{we18}) is equal to the fully separable state $\varrho .$ If some of
the terms in bracket of the left hand side of (\ref{we18}) are not equal
with each other, then we have $\rho =(1-\kappa )\rho _d+\kappa \varrho $,
with $0\leq \kappa <1$ and $\rho _d$ is a diagonal state in computational
basis thus fully separable. $\blacksquare $

\section{\textit{\ }The tripartite separability of noisy four qubit GHZ state
}

The four qubit GHZ state $\left| GHZ_4\right\rangle $ is a graph state
characterized by its four stabilizer generators $XZZZ,$ $ZXII,$ $ZIXI,$ $%
ZIIX.$ We may apply Hadamard transformations on all the qubits except the
first one, the generators then become $K_1=XXXX,$ $K_2=ZZII,$ $K_3=ZIZI,$ $%
K_4=ZIIZ.$ In Pauli matrix form, we have $\left| GHZ_4\right\rangle
\left\langle GHZ_4\right| =\frac 1{16}(IIII+ZZII+ZIZI+ZIIZ$ $+IZZI+IZIZ$ $%
+IIZZ$ $+ZZZZ$ $+$ $XXXX$ $-YYXX$ $-YXYX$ $-YXXY$ $-XYYX$ $-XYXY$ $-XXYY$ $%
+YYYY).$ The noisy GHZ state is $\rho =p\left| GHZ_4\right\rangle
\left\langle GHZ_4\right| +\frac{1-p}{16}I_{16}.$ The biseparability and
full separability of $\rho $ are known \cite{GuhneNJP}. For the
tri-separability, we find the matched witness with parameters $%
M_{3300}=M_{3030}$ $=M_{3003}=M_{0330}$ $=M_{0303}=M_{0033}=0,$ $M_{3333}=2,$
$M_{1111}=M_{2222}$ $=1,$ $M_{2211}=M_{2121}$ $=M_{2112}=M_{1221}$ $%
=M_{1212}=M_{1122}=-1.$ In the following, we will show that $M_{0000\text{ }%
} $is $-2$. The critical $p$ for tri-separable is $p=\frac 15.$ The state $%
\rho $ is a mixture of three part product states if $p\leq \frac 15$ and can
not be three partite separable for $p>\frac 15.$

\subsection{The necessary condition}

Consider the qubits $1,2,3,4$, we first classify the qubits into three parts
with the first two qubits in a part, the third and the fourth are the other
two parts. We denote the partition as $12|3|4.$ For the given $M_{klmn}$
except $M_{0000}$ in the main text, we have $-M_{0000}=\max f,$ where $%
f=2T_{33}z_3z_4+(T_{11}+T_{22})(x_3x_4+y_3y_4)+(T_{21}+T_{12})(y_3x_4+x_3y_4),
$ subject to $x_i^2+y_i^2+z_i^2=1$ $(i=3,4)$ and $T_{ij}=\left\langle \psi
\right| \sigma _i\otimes \sigma _j\left| \psi \right\rangle ,$ where $\left|
\psi \right\rangle =\alpha \left| 00\right\rangle +\beta \left|
01\right\rangle $ $+\gamma \left| 10\right\rangle $ $+\delta \left|
11\right\rangle $ is a two qubit pure state. Then $T_{33}=\left| \alpha
\right| ^2+\left| \delta \right| ^2-(\left| \beta \right| ^2+\left| \gamma
\right| ^2),$ $T_{11}+T_{22}=2(\beta \gamma ^{*}+\gamma \beta ^{*}),$ $%
T_{21}+T_{12}=2i(\alpha \delta ^{*}-\delta \alpha ^{*}).$ Denote $a=z_3z_4,$
$b=x_3x_4+y_3y_4,$ $c=y_3x_4+x_3y_4.$ The extremal values of $f$ are $a\pm
2c=2z_3z_4\pm 2(y_3x_4+x_3y_4)\leq 2$ and $-a\pm 2c$ $=-2z_3z_4$ $\pm
2(x_3x_4+y_3y_4)$ $\leq 2.$ The maximum of $f$ is $2.$ Thus $f\leq 2$ for $%
12|3|4$ partition. By the symmetry of the problem, we have $f\leq 2$ for all
$6$ kinds of partitions. Thus we have $M_{0000}=-2,$ the necessary condition
for tri-separablity is obtained as $p\leq \frac 15.$

\subsection{The sufficient condition}

The noisy four qubit $GHZ$ state with $p=\frac 15$ can be written as $\rho $
$=\frac 1{16}[\frac 15(IIII+XXXX+ZZII$ $-YYXX)$ $+\frac
15(IIII+YYYY-YXYX+IZIZ)$ $+\frac 15(IIII-YXXY-XXYY+ZIZI)+\frac
15(IIII-XYYX-XYXY+IIZZ)$ $+\frac 15(IIII+ZIIZ+IZZI+ZZZZ)].$ Each round
bracket in the above expression is tri-separable. The first four round
brackets are tri-separable in the following partitions respectively, $%
12|3|4, $ $13|2|4,$ $1|3|24,$ $1|2|34.$ The last one is fully separable. For
example, the first round bracket is $(IIII+XXXX+ZZII$ $-YYXX)=\frac
12(II+ZZ+XX-YY)(II+XX)$ $+\frac 12(II+ZZ-XX+YY)(II-XX).$ The components $%
(II\pm XX)$ $=\frac 12(I+X)(I\pm X)$ $+\frac 12(I-X)(I\mp X)$ is separable
for the third and the fourth qubits. The components $[II+ZZ\pm (XX-YY)]$ is
proportional to valid two qubit states for the first two qubits. Thus $%
IIII+XXXX+ZZII$ $-YYXX$ is tri-separable for partition $12|3|4.$

\section{The full separability of noisy four qubit cluster state}

The four qubit cluster state $\left| Cl_4\right\rangle $ is a graph state
characterized by its four stabilizer generators $K_1=XZII,$ $K_2=ZXZI,$ $%
K_3=IZXZ,$ $K_4=IIZX,$ where $X,Y,Z$ are the Pauli matrices. The noisy
cluster state is
\begin{equation}
\rho =p\left| Cl_4\right\rangle \left\langle Cl_4\right| +\frac{1-p}{16}%
I_{16}.  \label{wee1}
\end{equation}
The biseparability of the state is known \cite{GuhneNJP}. We will consider
the full separability of the state in this section and the three partite
separablilty in the next section. The cluster state can be written as $%
\left| Cl_4\right\rangle \left\langle Cl_4\right| =\frac
1{16}\prod_{j=1}^4(I+K_j)=\frac
1{16}\sum_{i_1,i_2,i_3,i_4=0}^1K_4^{i_4}K_3^{i_3}K_2^{i_2}K_1^{i_1}.$ In the
form of Pauli matrices, we have $\left| Cl_4\right\rangle \left\langle
Cl_4\right| =\frac 1{16}(IIII+XZII+ZXZI$ $+YYZI$ $+IZXZ$ $%
+XIXZ+ZYYZ-YXYZ+IIZX$ $+XZZX+ZXIX+YYIX+IZYY+XIYY-ZYXY+YXXY).$ The
characteristic function $R_{klmn}$ of $\rho $ has the values of $1$, $p$,$-p$
or $0.$ The number of nonzero $R_{klmn}$ is $16.$ The matched witness has $16
$ parameters $M_{klmn}.$

For the full separability, we find that the witness with the following
parameters is a matched witness, $M_{3130}=M_{0313}$ $=M_{3101}=M_{1013}$ $%
=M_{2230}=M_{0322}=$ $M_{2201}=M_{1022}=1,$ $M_{3223}=M_{1331}=M_{2112}=2,$ $%
M_{2123}=M_{3212}=-2,$ $M_{1300}=M_{0031}=0$. In the following we will show
that $M_{0000\text{ }}$is $-2$. Hence, the noise tolerance of full
separability for $\left| Cl_4\right\rangle $ is $p_{tol}=1-p$ with $p=\frac
19.$ The state $\rho $ is fully separable for $p\leq \frac 19$ and is
entangled for $p>\frac 19.$

\subsection{The necessary condition}

For the given $M_{klmn}$ except $M_{0000},$ we have $-M_{0000}=\max f,$
where $f=\{(z_1x_2+y_1y_2)(z_3+x_4)$ $+(z_2+x_1)(x_3z_4+y_3y_4)$ $%
+2(z_1y_2y_3z_4+x_1z_2z_3x_4+y_1x_2x_3y_4-y_1x_2y_3z_4-z_1y_2x_3y_4)\},$
subject to $x_i^2+y_i^2+z_i^2=1$ $(i=1,\ldots ,4).$ Using $%
x_4^2+y_4^2+z_4^2=1,$ the maximization over $(x_4,y_4,z_4)$ leads to $f\leq
f_1$ where
\[
f_1=b\cos \theta +\sqrt{(b+d\cos \theta )^2+e^2\sin ^2\theta }.
\]
With $b=$ $z_1x_2+y_1y_2,$ $d=2x_1z_2,$ $e=\sqrt{%
(z_2+x_1)^2+4(z_1y_2-y_1x_2)^2},$ and $z_3=\cos \theta $ is assumed. There
are two solutions for the maximization of $f_1$ over $\theta $. The first is
$\sin \theta =0,$thus $f_1=\pm b+\left| \pm b+d\right| \leq 2.$ The
inequality comes from $f_1=\pm 2b+d$ $=2[\pm \left( z_1x_2+y_1y_2\right)
+x_1z_2]$ $\leq 2$ when $\pm b+d\geq 0$ and $f_1=-d\leq 2$ when $\pm b+d<0.$
The second solution is $\cos \theta =\frac b{e-d}$ subject to $-1\leq \frac
b{e-d}\leq 1.$ Thus $f_1=\frac{b^2}{e-d}+e.$ A simple numeric calculation
shows that $f_1\leq 2.$ Hence $M_{0000}=-2.$ The necessary condition of full
separability is $p\leq \frac 19.$

\subsection{The sufficient condition}

The maximization of $f$ in above subsection hints the process of decomposing
a separable state into its explicit separable expression. If the maximal $%
f_1=2$ is achieved by one of the terms, say $2z_1y_2y_3z_4$, we have $%
2z_1y_2y_3z_4=2,$ The solutions of $z_1y_2y_3z_4=1$ are that $%
z_1,y_2,y_3,z_4 $ should be equal to $\pm 1$, the number of $-1$ should be
odd. Then there are $8$ solutions. Each solution corresponds to a product
state, for example $z_1=y_2=y_3=z_4=1$ corresponds to a product state
proportional to $(I+Z)(I+Y)(I+Y)(I+Z).$ Summing up all the $8$ product
states gives rise to unnormalized fully separable state $IIII+ZYYZ.$
Similarly, all the other solutions of $f=2$ can be utilized to obtained the
product states. Thus the mixture of the product states will compose the
noisy cluster state if the noise is under some threshold.

The noisy cluster state with $p=\frac 19$ can be written as $\rho =\frac
1{16}[\frac 19(IIII+ZYYZ)$ $+\frac 19(IIII+YXXY)$ $+\frac
19(IIII-YXYZ)+\frac 19(IIII-ZYXZ)+\frac 19(IIII+ZXZI+ZXIX+IIZX)$ $+\frac
19(IIII+YYZI+YYIX+IIZX)$ $+\frac 19(IIII+IZXZ+XIXZ+XZII)$ $+\frac
19(IIII+IZYY+XIYY+XZII)$ $+\frac 19(IIII-IIZX-XZII+XZZX)].$

Each round bracket in the above expression is fully separable.

\section{Noise tolerance of four qubit cluster state in three parties}

For the tri-separability, we find the witness to be $W=cIIII-Q,$ where
\begin{eqnarray}
Q &=&-XZII-IIZX+3XZZX  \nonumber \\
&&+ZXZI+XIYY+3YXXY  \nonumber \\
&&+YYZI+XIXZ+3ZYYZ  \nonumber \\
&&+ZXIX+IZXZ-3ZYXY  \nonumber \\
&&+YYIX+IZYY-3ZYXY,  \label{wee2}
\end{eqnarray}
and $c=5.$ The entanglement is detected if $tr(\rho W)<0,$ which leads to $%
p>\frac 5{21}.$ Hence the state $\rho $ is a mixture of three part product
states if $p\leq \frac 5{21}$ and can not be three partite separable for $%
p>\frac 5{21}.$

\subsection{The necessary condition}

For all four qubit tripartite separable state $\rho _{3sep},$ we should have
$tr(\rho _{3sep}W)\geq 0,$ that is
\begin{equation}
c=\max_{\rho _{3sep}}tr(\rho _{3sep}Q).  \label{wee3}
\end{equation}

Consider the qubits $1,2,3,4$, we first classify the qubits into three parts
with the first two qubits in a part, the third and the fourth are the other
two parts. We denote the partition as $12|3|4.$ The tripartite separable
state for this partition is $\rho _{12|3|4}=\sum_ip_i\rho _{12}^{(i)}\otimes
\rho _3^{(i)}\otimes \rho _4^{(i)},$ where $p_i$ form a probability
distribution. Without loss of generality, we consider the states in each
party to be pure. The tripartite separable state is a mixture of pure
product state $\left| \psi _{12}\right\rangle \left| \psi _3\right\rangle
\left| \psi _4\right\rangle .$ Then for the partition $12|3|4,$ we have $%
c=\max_{\left| \psi _{12}\right\rangle \left| \psi _3\right\rangle \left|
\psi _4\right\rangle }\left\langle \psi _{12}\right| \left\langle \psi
_3\right| \left\langle \psi _4\right| Q\left| \psi _{12}\right\rangle \left|
\psi _3\right\rangle \left| \psi _4\right\rangle .$ Then $c=\max_{\left|
\psi _{12}\right\rangle }\left\langle \psi _{12}\right| \mathcal{M}\left|
\psi _{12}\right\rangle ,$ where the matrix $\mathcal{M}=$ $\left\langle
\psi _3\right| \left\langle \psi _4\right| Q\left| \psi _3\right\rangle
\left| \psi _4\right\rangle $ is the partial trace of $Q$ over the third and
the fourth qubits, respectively. Applying the Hadamard transform $H_2$ on
the first qubit, we obtain the matrix $\mathcal{M}^{\prime }=(H_2\otimes I)%
\mathcal{M}(H_2\otimes I),$ the eigenvalues do not change since the Hadamard
transform is unitary. Hence $c$ is equal to the largest eigenvalue of $%
\mathcal{M}^{\prime }.$ Denote the Bloch vectors of $\left| \psi
_i\right\rangle $ as $(x_i,y_i,z_i)$ with $i=1,...,4$, it follows that, $%
\mathcal{M}^{\prime }=(3z_3x_4-1)ZZ-z_3x_4II+(x_3z_4+y_3y_4)(ZI+IZ)$ $%
+(z_3+x_4)(XX-YY)$ $+3(y_3z_4-x_3y_4)(XY+YX).$ The eigenvalues of $\mathcal{M%
}^{\prime }$ are $\lambda _{1,2}=2z_3x_4-1\pm 2\sqrt{%
(x_3z_4+y_3y_4)^2+(z_3+x_4)^2+9(y_3z_4-x_3y_4)^2},$ and $\lambda
_{3,4}=1-4z_3x_4.$ The maximum of $\lambda _1$ is $5$ and it is achieved
when
\begin{equation}
z_3=x_4=0,y_3z_4-x_3y_4=\pm 1;  \label{wee4}
\end{equation}
or
\begin{equation}
z_3=x_4=\pm 1.  \label{wee5}
\end{equation}
The maximums of $\lambda _3$ and $\lambda _4$ are $5$ and they are achieved
when
\begin{equation}
z_3=-x_4=\pm 1.  \label{wee6}
\end{equation}
Hence the eigenvalues of $\mathcal{M}^{\prime }$ is tight upper bounded by $%
5.$ We thus arrives $c=5$ for the partition $12|3|4.$

Then we consider the partition $13|2|4,$ the first and the third qubits are
in a party, the other two qubits are in the other two parties, respectively.
We have $c=\max_{\left| \psi _{13}\right\rangle \left| \psi _2\right\rangle
\left| \psi _4\right\rangle }\left\langle \psi _{13}\right| \left\langle
\psi _2\right| \left\langle \psi _4\right| Q\left| \psi _{13}\right\rangle
\left| \psi _2\right\rangle \left| \psi _4\right\rangle =\max_{\left| \psi
_{13}\right\rangle }\left\langle \psi _{13}\right| \mathcal{M}\left| \psi
_{13}\right\rangle ,$ with matrix $\mathcal{M}=$ $\left\langle \psi
_2\right| \left\langle \psi _4\right| Q\left| \psi _2\right\rangle \left|
\psi _4\right\rangle .$ Applying Hadamard transform to the first qubit and
eliminating the phase factors of the matrix entries with unitary
transformation, then the matrix $\mathcal{M}$ is transformed to $\mathcal{M}%
^{\prime }=U(H_2\otimes I)\mathcal{M}(H_2\otimes I)U^{\dagger }$, namely,
\begin{equation}
\mathcal{M}^{\prime }=\left[
\begin{array}{llll}
K_1 & \alpha _4z_{2+} & \alpha _2x_{4+} & -3\alpha _2\alpha _4 \\
\alpha _4z_{2+} & K_2 & 3\alpha _2\alpha _4 & \alpha _2x_{4-} \\
\alpha _2x_{4+} & 3\alpha _2\alpha _4 & K_3 & \alpha _4z_{2-} \\
-3\alpha _2\alpha _4 & \alpha _2x_{4-} & \alpha _4z_{2-} & K_4
\end{array}
\right] .  \label{wee7}
\end{equation}
Where $z_{2\pm }=z_2\pm 1,$ $x_{4\pm }=x_4\pm 1,$ $\alpha _2=\sqrt{1-z_2^2},$
$\alpha _4=\sqrt{1-x_4^2},$ $K_{1,3}=\pm 3z_2x_4\mp z_2-x_4,$ $K_{2,4}=\mp
3z_2x_4\mp z_2+x_4.$ The unitary transformation for eliminating the matrix
entry phase factors is $U=diag\{1,e^{i\theta _4},e^{-i\theta _2},e^{i(\theta
_4-\theta _2)}\}$, where $\theta _2=\tan ^{-1}(y_2/x_2),\theta _4=\tan
^{-1}(y_4/z_4).$ Thus $.$ Let $t=x_4^2\alpha _2^2+z_2^2\alpha _4^2,$ the
eigenequation is $\lambda ^4-2(11-8t)\lambda ^2+8(4t-3)\lambda +3(15-16t)=0,$
which can be factorized to $\left( \lambda ^2+2\lambda -3\right) (\lambda
^2-2\lambda +16t-15)=0,$ the maximal eigenvalue is $\lambda _m=1+4\sqrt{1-t}%
\leq 5.$ Hence we have $c=5$ for the partition $13|2|4.$ The maximal
eigenvalue is achieved when $t=0$, namely
\begin{equation}
x_4=\pm 1,\text{ }z_2=\pm 1;  \label{wee8}
\end{equation}
or
\begin{equation}
x_4=0,\text{ }z_2=0.  \label{wee9}
\end{equation}

For the partition $14|2|3,$ we have $c=\max_{\left| \psi _{14}\right\rangle
}\left\langle \psi _{14}\right| \mathcal{M}\left| \psi _{14}\right\rangle ,$
with matrix $\mathcal{M}=$ $\left\langle \psi _2\right| \left\langle \psi
_3\right| Q\left| \psi _2\right\rangle \left| \psi _3\right\rangle .$ We can
transform the matrix $\mathcal{M}$ to $\mathcal{M}^{\prime }=U(H_2\otimes
H_2)\mathcal{M}(H_2\otimes H_2)U^{\dagger },$ with $U=diag\{1,e^{-i\theta
_3},e^{-i\theta _2},e^{-i(\theta _3+\theta _2)}\}$, where $\theta _3=\tan
^{-1}(y_3/x_3).$ The $\mathcal{M}^{\prime }$ has the same form as in (\ref
{wee7}) with $z_3,$ $x_4,$ $\alpha _3,\alpha _4$ being substituted by $%
z_2,z_3,\alpha _2,\alpha _3,$ respectively, where $\alpha _2=\sqrt{1-z_2^2}$%
. The maximal eigenvalue of $\mathcal{M}^{\prime }$ is $5,$ hence $c=5,$
which is achieved at
\begin{equation}
z_2=\pm 1,\text{ }z_3=\pm 1;  \label{wee10}
\end{equation}
or
\begin{equation}
z_2=0,\text{ }z_3=0.  \label{wee11}
\end{equation}

For the partition $1|23|4,$ we have $c=\max_{\left| \psi _{23}\right\rangle
}\left\langle \psi _{23}\right| \mathcal{M}\left| \psi _{23}\right\rangle ,$
with matrix $\mathcal{M}=$ $\left\langle \psi _1\right| \left\langle \psi
_4\right| Q\left| \psi _1\right\rangle \left| \psi _4\right\rangle .$ Using
unitary transform $U=diag\{1,e^{i\theta _4},e^{i\theta _1},e^{i(\theta
_1+\theta _4)}\},$ with $\theta _1=\tan ^{-1}(y_1/z_1),$ we obtain $\mathcal{%
M}^{\prime }$ having the same form as in (\ref{wee7}) with $z_3,$ $\alpha _3$
being substituted by $x_1,\alpha _1,$ respectively, where $\alpha _1=\sqrt{%
1-x_1^2}$. The maximal eigenvalue of $\mathcal{M}^{\prime }$ is $5,$ hence $%
c=5,$ which is achieved at
\begin{equation}
x_1=\pm 1,\text{ }x_4=\pm 1;  \label{wee12}
\end{equation}
or
\begin{equation}
x_1=0,\text{ }x_4=0.  \label{wee13}
\end{equation}

Since the witness is symmetric under exchange of the first qubit with the
fourth, the second with the third, it follows that $c=5$ for the partitions $%
1|2|34$ and $1|4|23.$ We have proven that $c=5$ for all six partitions.

\subsection{The sufficient condition}

For $p=\frac 5{21},$ we will prove explicitly that the noisy cluster state (%
\ref{wee1}) is tripartite separable. Let consider partition $1|23|4,$ the
maximal eigenvalue of $\mathcal{M}$ is achieved for the conditions (\ref
{wee12}) or (\ref{wee13}). The condition $x_1=1,$ $x_4=1\ $corresponds to
the state $\frac 12(I+X)$ for the first qubit and $\frac 12(I+X)$ for the
fourth qubit. The $x_1=1,$ $x_4=1$ also leads to a diagonal $\mathcal{M}$.
We have $\mathcal{M}=diag\{1,-3,-3,5\}.$ Hence the maximal eigenvalue is
achieved by $\left| \psi _{23}\right\rangle =\left| 11\right\rangle $. Thus
the pure state that achieves $c=5$ is $\frac 14(I+X)\left| 11\right\rangle
\left\langle 11\right| (I+X)$ $=\frac 1{16}(I+X)(I-Z)(I-Z)(I+X).$ The other
three cases of (\ref{wee12}) lead to three similar separable states. Average
on all these four state we arrive at the separable state
\begin{equation}
\rho _0=\frac 1{16}(IIII-XZII-IIZX+XZZX).  \label{wee14}
\end{equation}
The condition (\ref{wee13}) corresponds to states $\frac 12(I+\sin \theta
_iY+\cos \theta _iZ)$ for the first ($i=1$) and the fourth ($i=4$) qubits.
At this condition the eigenvector corresponds to the maximal eigenvalues is $%
\left| \psi _{23}\right\rangle =\frac 14(\left| 00\right\rangle +e^{i\theta
_4}\left| 01\right\rangle +e^{i\theta _1}\left| 10\right\rangle -e^{i\theta
_1+i\theta _4}\left| 11\right\rangle ).$ We may write $\left| \psi
_{23}\right\rangle \left\langle \psi _{23}\right| =\frac 14(II$ $+\cos
\theta _1XZ+\sin \theta _1\sin \theta _4XX+\sin \theta _1YZ+\cos \theta
_1\cos \theta _4YY$ $+\cos \theta _4ZX-\cos \theta _1\sin \theta _4YX+\sin
\theta _4ZY-\sin \theta _1\cos \theta _4XY$ $).$ Denote the tripartite
separable state as $\varrho _1(\theta _1,\theta _4)=\frac 14(I+\sin \theta
_1Y+\cos \theta _1Z)$ $\left| \psi _{23}\right\rangle \left\langle \psi
_{23}\right| (I+\sin \theta _4Y+\cos \theta _4Z).$ Let $\overline{\varrho }%
_1=\frac 1{16}\sum_{j,k=0}^3\varrho _1(\frac{(2j+1)\pi }4,\frac{(2k+1)\pi }%
4) $. Thus we have the tripartite separable state
\begin{equation}
\overline{\varrho }_1=\xi +\frac 1{32}(IZXZ+IZYY+YYZI+ZXZI),  \label{wee15}
\end{equation}
where $\xi =\frac 1{16}IIII+\frac 1{64}(YXXY-YXYZ+ZYYZ-ZYXY).$ Similarly, we
have the tripartite separable states
\begin{equation}
\overline{\varrho }_2=\xi +\frac 1{32}(XIXZ+XIYY+YYIX+ZXIX),  \label{wee16}
\end{equation}
\begin{equation}
\overline{\varrho }_3=\xi +\frac 1{32}(ZXZI+XIYY+YYZI+XIXZ),  \label{wee17}
\end{equation}
\begin{equation}
\overline{\varrho }_4=\xi +\frac 1{32}(IZXZ+YYIX+IZYY+ZXIX),  \label{wee18}
\end{equation}
for partitions $14|2|3,13|2|4,1|3|24,$ respectively. We have a tripartite
separable state
\begin{equation}
\rho _1=\frac 12(\overline{\varrho }_1+\overline{\varrho }_2)=\frac 12(%
\overline{\varrho }_3+\overline{\varrho }_4).  \label{wee19}
\end{equation}

For the partition $12|3|4,$ the maximal eigenvalue of $\mathcal{M}^{\prime }$
is achieved at the condition of either (\ref{wee4}) or (\ref{wee5}), or (\ref
{wee6}). When $z_3=x_4=1,$ the third and the fourth qubits are in the states
$\frac 12(I+Z)$ and $\frac 12(I+X)$, respectively. The $\mathcal{M}^{\prime
} $ is reduced to $2ZZ-II+2(XX-YY)$ with eigenfunction $\frac 1{\sqrt{2}%
}(\left| 00\right\rangle +\left| 11\right\rangle )$ for its largest
eigenvalue $\lambda _1=5.$ The corresponding eigenfunction for $\mathcal{M}$
is $\left| \psi _{12}\right\rangle =\frac 12(\left| 00\right\rangle +\left|
01\right\rangle +\left| 10\right\rangle -\left| 11\right\rangle ).$ Hence
the tripartite separable state that achieves the condition $c=5$ is $\frac
14\left| \psi _{12}\right\rangle \left\langle \psi _{12}\right| (I+Z)(I+X)$ $%
=\frac 1{16}(II+XZ+ZX+YY)(I+Z)(I+X).$ Similarly, the tripartite separable
state that achieves the condition $c=5$ is $\frac
1{16}(II+XZ-ZX-YY)(I-Z)(I-X)$ for the case $z_3=x_4=-1.$ The average of
these two states gives rise to the tripartite separable state $\varrho
_5=\frac 1{16}(IIII+XZII+IIZX$ $+XZZX$ $+ZXZI+YYZI+ZXIX+YYIX).$ By the
symmetry, we have the tripartite separable state $\varrho _6=\frac
1{16}(IIII+XZII+IIZX+XZZX+IZXZ+IZYY+XIXZ+XIYY)$ for the partition $1|2|34.$
We thus have a tripartite state
\begin{equation}
\rho _2=\frac 12(\varrho _5+\varrho _6).  \label{wee20}
\end{equation}
For the case of (\ref{wee4}), $\mathcal{M}^{\prime }$ is reduced to $-ZZ\pm
3(XY+YX)$, where the sign $\pm $ are for the cases $y_3z_4-x_3y_4=\pm 1$,
respectively. The eigenvector for the largest eigenvalue $\lambda _1=5$ is $%
\frac 1{\sqrt{2}}(\left| 00\right\rangle \pm i\left| 11\right\rangle ).$ The
corresponding eigenvector of $\mathcal{M}$ is $\left| \psi
_{12}\right\rangle =\frac 12(\left| 00\right\rangle \pm i\left|
01\right\rangle $ $+\left| 10\right\rangle \mp i\left| 11\right\rangle ).$
Hence $\left| \psi _{12}\right\rangle \left\langle \psi _{12}\right| =\frac
14[II+XZ\pm (YX-ZY)].$ The third and the fourth qubits are $\frac 12(I+\cos
\theta _3X+\sin \theta _3Y)$ and $\frac 12(I+\sin \theta _4Y+\cos \theta
_4Z) $ with $\theta _4=\theta _3\pm \frac \pi 2,$ respectively. The product
state of the third and the fourth qubits then is $\frac 14(I+\cos \theta
_3X+\sin \theta _3Y)$ $(I\mp \sin \theta _3Z\pm \cos \theta _3Y).$ The
average on $\theta _3=\frac \pi 4,\frac{3\pi }4,\frac{5\pi }4\frac{7\pi }4$
is the state of $\frac 14[II\pm \frac 12(XY-YZ)].$ The tripartite separable
states are $\frac 1{16}[II+XZ\pm (YX-ZY)][II\pm \frac 12(XY-YZ)].$ Averaging
on the $\pm $ states gives rise to a tripartite separable state $\varrho
_7=\frac 1{16}(II+XZ)II$ $+\frac 1{32}(YX-ZY)(XY-YZ).$ By the symmetry, for
the partition $1|2|34$, we have tripartite separable state $\varrho _8=\frac
1{16}II(II+ZX)$ $+\frac 1{32}(YX-ZY)(XY-YZ).$ Thus we have a tripartite
separable state
\begin{equation}
\rho _3=\frac 12(\varrho _7+\varrho _8).  \label{wee21}
\end{equation}
At last, we can compose the tripartite separable state as
\begin{eqnarray}
\rho &=&\frac 1{21}(\rho _0+12\rho _1+4\rho _2+4\rho _3)  \nonumber \\
&=&\frac 5{21}\left| Cl_4\right\rangle \left\langle Cl_4\right| +\frac
1{21}IIII.  \label{wee22}
\end{eqnarray}

\section{Conclusion}

We have utilized the characteristic coeffients (variables) of witness
operator to investigate the multipartite separability of multipartite
quantum states. The necessary condition of separability can be obtained for
any given set of characteristic variables as far as the algebraic
maximization can be worked out. The sufficient criterion is obtained by
matching the characteristic variables to the given quantum state whose
separability is under researched. We use the three qubit X shaped state to
illustrate the process of finding the necessary and sufficient criterion of
full separability with our method. New results are the necessary and
sufficient conditions for tripartite separability and full separability of
four qubit cluster state in white noise, the necessary and sufficient
conditions for the tripartite separability of four qubit GHZ state in white
noise. The noise tolerances of the tripartite separability and full
separability of four qubit cluster state are $\frac{16}{21}$ and $\frac 89,$
respectively. The noise tolerance of the tripartite separability of four
qubit GHZ is $\frac 45.$ These conditions are necessary and sufficient. We
also explicitly construct the separable states for these four qubit noisy
states with given parties. The matched witness method is suitble in finding
the multipartite separable criterion for quatum states with less
characteristic variables.

\section*{Acknowledgment}

Supported by the National Natural Science Foundation of China (Grant Nos.
11375152) and (partially) supported by National Basic Research Program of
China (Grant No. 2014CB921203) are gratefully acknowledged.

\section*{Appendix}

\textit{A. Proof of Lemma 1. }

Proof: It is not difficult to eliminate two of the angles in $g(\mathbf{%
\varphi })$, say, $\varphi _1$ and $\varphi _2$ by maximization. We have
\begin{equation}
g(\mathbf{\varphi })\leq \sqrt{A}+\sqrt{B},  \label{we10}
\end{equation}
with $A=\delta ^2+$ $\gamma ^2-2\delta \gamma \cos 2\varphi _3,$ $B=\alpha
^2+$ $\beta ^2+2\alpha \beta \cos 2\varphi _3.$ The solutions of $\frac{%
dg_1(\varphi _3)}{d\varphi _3}=0$ are (i) $\sin 2\varphi _3=0$, it gives
rise to the second line of (\ref{we9}), and (ii)
\begin{equation}
\cos 2\varphi _3=\frac{\alpha ^2\beta ^2(\delta ^2+\gamma ^2)-\gamma
^2\delta ^2(\alpha ^2+\beta ^2)}{2\alpha \beta \gamma \delta (\gamma \delta
+\alpha \beta )},  \label{we11}
\end{equation}
it gives rise to the first line of (\ref{we9}). The condition for the
existence of the solution (ii) is $\left| \cos 2\varphi _3\right| \leq 1,$
which leads to $q\geq 0.$ The condition $\delta \alpha \beta \gamma >0\ $%
comes from $q_0^2=(\delta \alpha +\beta \gamma )(\delta \beta +\alpha \gamma
)(\delta \gamma +\alpha \beta )-\delta \alpha \beta \gamma M_{111}^2\geq 0.$
Thus when $\delta \alpha \beta \gamma >0,$ the first line of (\ref{we9}) is
larger than the second line.

\textit{B. Proof of (\ref{we14a}) }

We start from Eq. (\ref{we12}). The aim is to maximize $F_2(\theta _1,\theta
_2)$ when $a=M_1c_1,$ $b=M_2+M_3c_1,$ $c=M_3+M_2c_1,$ $d=M_1+M_0c_1,$ $%
e=g_ms_1$ with $c_1=\cos \theta _1,s_1=\sin \theta _1$ and $g_m$ is related
with $M_i$ as in the main text. The maximization of $F_2$ with respect to $%
\theta _2$ leads to two solutions. The first solution is $\sin \theta _2=0,$
thus for $F_3(\theta _1)=\max_{\theta _2}F_2(\theta _1,\theta _2),$ we have
\begin{equation}
F_3(\theta _1)=a+b+\left| c+d\right| ,  \label{Wave1}
\end{equation}
which is a linear function of $c_1.$ The maximal value is $F_4=-M_0$ when $%
-M_0$ is positive and large enough. The second solution is
\begin{equation}
\cos \theta _2=\frac 1{e^2-d^2}\left( cd+b\sqrt{\frac{e^2(c^2+e^2-d^2)}{%
(b^2+e^2-d^2)}}\right) .  \label{Wave2}
\end{equation}
Notice that the second solution does not exist if the absolute of right hand
side of (\ref{Wave2}) exceeds $1.$ The maximum of $F_2(\theta _1,\theta _2)$
with respect to $\theta _2$ is
\begin{eqnarray}
F_3(\theta _1) &=&a+\frac 1{e^2-d^2}(bcd+sign(b^2+e^2-d^2)  \nonumber \\
&&\times \sqrt{e^2(c^2+e^2-d^2)(b^2+e^2-d^2)}).  \label{Wave3}
\end{eqnarray}
The equation can be rewritten as
\begin{equation}
e^4+(c^2+b^2-d^2-h^2)e^2+(hd-bc)^2=0.  \label{Wave4}
\end{equation}
where $h=a-F_3(\theta _1).$ Suppose $\left. F_3(\theta _1)\right| _{\theta
_1=\theta _0}$ $=-M_{0\text{ }}$for some $\theta _0,$ then the solution of (%
\ref{Wave4}) is $\left. \cos \theta _1\right| _{\theta _1=\theta _0}=\frac{%
-A+\sqrt{A^2-B^2}}B,$ where $A=M_0^2+M_1^2-M_2^2-M_3^2,$ $%
B=2(M_0M_1-M_2M_3). $ Thus $-M_{0\text{ }}$ is an achievable value of
function $F_3(\theta _1).$ The derivative of Eq. (\ref{Wave4}) with respect
to $x=\cos \theta _1$ at gives rise to $\left. \frac{dh}{dx}\right|
_{x=x_0}=M_1$ and $\left. \frac{d^2h}{dx^2}\right| _{x=x_0}=-\left. \frac{%
A^2-B^2}{e^2h}\right| _{\theta =\theta _0},$ where $x_0=\cos \theta _0.$
Hence we arrive at
\begin{eqnarray}
\left. \frac{dF_3}{dx}\right| _{x=x_0} &=&0,  \label{Wave5} \\
\left. \frac{d^2F_3}{dx^2}\right| _{x=x_0} &=&\left. \frac{A^2-B^2}{e^2h}%
\right| _{\theta =\theta _0}<0.  \label{Wave6}
\end{eqnarray}
The inequality comes from the fact that $h(\theta _0)=M_0+M_1\cos \theta _0$
$<0$ if we choose $\left| M_0\right| >\left| M_1\right| $ and $M_0$ is
negative. In order to make each of $m_i$ ($i=0,\ldots ,3$) positive, we have
to choose $M_0$ with such a property. Combining all of the solutions
together, we conclude that the maximum of $F_2(\theta _1,\theta _2)$ at case
(ii) is $-M_0$ for sufficiently large and positive $-M_0.$

\end{document}